# Effect of Ka-band Microwave on the spin dynamics of electrons in a GaAs/Al$_{0.35}$Ga$_{0.65}$As heterostructure

Haihui Luo, Xuan Qian, Xiaofang Gu, Yang Ji[*]

*SKLSM, Institute of Semiconductors, Chinese Academy of Sciences,*

*Beijing 100083, People's Republic of China*

V. Umansky

*Braun Center for Submicron Research, Department of Condensed Matter Physics,*

*Weizmann Institute of Science, Rehovot 76100, Israel*

We report experimental results of the effect of Ka-band microwave on the spin dynamics of electrons in a 2D electron system in a GaAs/Al$_{0.35}$Ga$_{0.65}$As heterostructure, via time-resolved Kerr rotation measurements. While the microwave reduces the transverse spin lifetime of the bulk GaAs when its frequency is close to the Zeeman splitting of the electrons in the magnetic field, it significantly increases that of electrons in the 2D electron system, from 745 ps to 1213 ps. Such a microwave-enhanced spin lifetime is ascribed to the microwave-induced electron scattering which leads to a "motional narrowing" of spins via D'yakonov-Perel' mechanism.

One of the most important issues in the ever-expanding field of semiconductor spintronics is the coherent manipulation of electron spins, which is critical to future information processing techniques (quantum computation in particular).[1-3] Being crucial for quantum information application,[1] a long spin lifetime of electrons may be achieved by proper doping,[4] gate-voltage modulation,[5] heterostructure geometry,[6] suitable strain,[7] and/or an external magnetic field.[8] The coupling of spin nondegenerate systems and microwave photons has long been utilized to study spin dynamics in various materials, such as magnetic resonance and microwave-induced spin-flip in semiconductor two-dimensional electron systems (2DES),[9,10] spin states initialization and detection in nitrogen vacancy centers in diamond,[11] dynamical control over the molecular spin states[12] and the spin oscillations in ultracold atomic gases[13]. However, the microwave effect on the spin lifetime of electrons in semiconductor heterostructures has been rarely reported. Here we report on time-resolved Kerr rotation (TRKR) measurement of the effect of the Ka-band microwave on the spin dynamics of electrons in a 2DES in a GaAs/Al$_{0.35}$Ga$_{0.65}$As heterostructure in a magnetic field. We found that the transverse spin lifetime ($T_2^*$) of electrons in the 2DES can be significantly increased from 745 ps to 1213 ps, when the microwave frequency is close to the Zeeman splitting of the electrons in the magnetic field. This work provides a new method to prolong the spin lifetime in quantum-confined systems, which may be useful for future quantum information application.

Our sample is a high-quality, single-side modulation-doped (n=$3.1\times10^{18}$ cm$^{-3}$) GaAs/Al$_{0.35}$Ga$_{0.65}$As heterostructure with a 2DES mobility of $\mu$=$3.2\times10^6$ cm$^2$/Vs and an electron density of $n_{2D}$=$9.6\times10^{10}$ cm$^{-2}$ at $T$=4.2 K.[14] Optical measurements were performed in a magneto-optical cryostat with a superconducting split-coil magnetic field up to 10 T and with tunable temperature from 1.5 to 300 K. A special sample-rod with two rectangular wave-guides for 8mm microwave was designed to radiate the Ka-band (with frequency ranging from 26–40 GHz[15]) microwave onto the sample. A microwave generator with a tunable frequency of 33.5−36.5 GHz and a maximal power output of 50 mW was used. The microwave power available on the sample is about 5 mW, with its alternating electronic field parallel to the sample surface. For dynamical measurements of spin-polarized states, we performed Kerr spectra (KS) and TRKR measurement in a pump-probe configuration at low temperature within an in-plane magnetic field (the Voigt geometry). The sample was excited with laser beams from a mode-locked Ti: sapphire pulse laser (Coherent 900D) which has a temporal duration of 150 fs and a repetition rate of 76 MHz. The pump and probe beams, which had the same wavelength, were focused to a spot of ~100 μm in diameter, with constant powers of 5 mW and 0.5 mW, respectively. The KS was measured at a fixed probe delay of 30 ps by scanning the wavelengths of the laser beams. For lock-in detection of the TRKR, two different modulations were used: one is the circular polarization modulation of the pump beam with a photoelastic modulator at 50 kHz and the other is the output modulation of the microwave power at 513 Hz.

Figure 1 shows a KS trace measured at a fixed time delay of 30 ps at $T$=1.5 K and $B$=5.6 T. The spectral curve have a peak and a dip in the excited energy ($E_p$) regime, which can be fitted with two symmetric Lorentz curves with centers at $E_p$=1.516 ev and $E_p$=1.522 ev, respectively, and full width half maximum of 6 mev and 16 mev, respectively. Note that these two Lorentz fits have opposite signs: the peak at low energy side is attributed to heavy hole – electron excitation, whereas the dip at high energy side is ascribed to light hole – electron excitation, with a reasonable energy gap of 6 mev between light-holes and heavy-holes in our sample.[16]

Figure 2a shows two TRKR traces taken at $B$=5.6 T with and without the microwave with a frequency of 35 GHz, indicating a dramatic change, induced by the microwave, of the envelope of the TRKR traces, which contains oscillations with different Larmor precession frequencies. The spin precessions are most clearly seen from the corresponding fast Fourier transform spectra (FFTS) in Fig. 2b of the TRKR data in Fig. 2a: there are two peaks and a broad band in the spin precession frequencies, with or without the microwave. The origin of the broadened spin precession frequency, which is of no importance to the following discussion and our main conclusions, is not clear yet but we assign it to exciton. As for the two peaks, we attribute the low frequency part of $\nu$=33.98 GHz and high frequency part of $\nu$=34.53 GHz to the 2DES and the bulk GaAs, respectively, as in our previous work.[14] Note that the microwave frequency of 35 GHz is close to the Zeeman splitting of the electrons in the magnetic field.

Figure 2b shows that the Ka-band microwave irradiation can significantly change the spectral weight of the three components: it increases the 2DES part and reduces the GaAs part and the exciton part. To see the transeverse spin lifetime and the initial Kerr rotation amplitude of these components (which may be retrieved from their distinct g-factor), we fit the spin precession data with the well established fitting method in which each exponentially decaying oscillatory function represents one spin precession.[17,18] Our fitting contains three cosine functions with exponentially-decay envelope that correspond to the two peaks and the broad band shown in Fig. 2b. The fitting curves, as shown by the solid curves in Fig. 2a, agree very well with the experimental TRKR traces and allow us to extract the Kerr signals arising from the 2DES, the bulk GaAs and the exciton through their distinct spin precession frequencies. The decompositions of TRKR signals with and without microwave irradiation are plotted in Fig. 2c and Fig. 2d, respectively.

Comparing the results in Fig. 2c with those in Fig. 2d, we found that when the sample is illuminated with the microwave, the transverse spin lifetime $T_2^*$ of electrons in the 2DES is significantly increased from 745 ps to 1213 ps, while that in the bulk GaAs is reduced from 422 ps to 274 ps. The response of the TRKR trace at the $E_p$=1.522 ev excitation to the microwave irradiation is similar to that of $E_p$=1.516 ev excitation, say, a significant increase of 2DES spin lifetime and a slight decrease of the electrons in GaAs and the excitons (thus not shown here).

We also modulated the microwave power output and measured the microwave-induced TRKR change, namely, the difference between TRKR traces with and without microwave, ΔTRKR, via lock-in detection. The ΔTRKR traces of two different excitations are shown in Fig. 3. The ΔTRKR decay envelope of the $E_p$=1.516 ev excitation contains two processes with different frequencies which correspond to the 2DES and GaAs spin precession frequencies, respectively, as revealed by its FFT result. As for the ΔTRKR trace of the $E_p$=1.522 ev excitation, microwave irradiation mainly affects spin precessions of the 2DES and the exciton, leading to a more obvious beating. In other words, the ΔTRKR results provide a direct view of the microwave-induced spin dynamics change and a solid support for the experimental results shown in Fig. 2.

The microwave irradiation has two main effects on the $T_2^*$ of electrons: on the one hand, it induces spin-flip processes of electrons, both in the resonant and off-resonant cases, as described by a spin-flip time $T_{sf}$; on the other hand, it affects the phase smearing of electron spins, which may be described by a spin dephasing time $T_{22}^*$. This may be expressed in the following formula:

$$\frac{1}{T_2^*} = \frac{1}{T_{sf}} + \frac{1}{T_{22}^*} \qquad (1)$$

Experimental results have demonstrated the spin relaxation rate ($\tau^{-1}$)−momentum relaxation time ($\tau_p(\mathbf{k})$) correlation in the high-mobility low-density 2DES, $\tau^{-1}=\langle\mathbf{\Omega}(\mathbf{K})^2\rangle\tau_p(\mathbf{k})$, in which the spin lifetime prolongs with the increasing electron-electron and electron- acoustic phonon scattering rate.[19] $T_{22}^*$ of the 2DES increases with the microwave irradiation because of the microwave-assisted electron-electron and electron-acoustic phonon scattering and the microwave induced heating. Another premise for the prolonged spin lifetime $T_2^*$ in 2DES is that the microwave-induced spin-flip time $T_{sf}$ is still long. Tough we can not obtain the exact value of $T_{sf}$, a very long microwave-induced $T_{\text{sf}}$ can be expected in low dimensional semiconductor materials.[9] However, there is a weak dependence of the electron $T_2^*$ on the electron scattering mechanisms in GaAs,[14] manifesting an insensitive $T_{22}^*$ to the presence of microwave irradiation, so $T_2^*$ decreases due to the microwave-induced spin-flip. It is worth pointing out that there is a weak microwave frequency dependence of the spin dynamics in the microwave frequency regime from 33.5 GHz to 36.5 GHz which is close the Zeeman splitting of the electrons of the 2DES and the GaAs in the magnetic field. It may be ascribed to the fact that there is not intense electron spin resonance in our sample,[20] since the electron densities in both of the 2DES and the GaAs are fairly low.

In summary, we have experimentally studied the effect of Ka-band microwave on the spin dynamics of electrons in a GaAs/Al$_{0.35}$Ga$_{0.65}$As heterostructure in a magnetic field, via TRKR measurements. We found that the $T_2^*$ of electrons in the 2DES can be significantly increased from 745 ps to 1213 ps, when the microwave frequency is close to the Zeeman splitting of the electrons in the magnetic field. This phenomenon can be interpreted by microwave-induced electron scattering which leads to "motional narrowing" of spins via D'yakonov-Perel' mechanism. Our work provides a new method to prolong the spin lifetime in quantum-confined systems, which may be useful for future quantum information application.

We thank M.W. Wu and X.Z. Ruan for helpful discussions. This work was supported by the NSFC under Grants No. 10425419, National Basic Research Program of China (No. 2007CB924900, No. 2009CB929301) and the Knowledge Innovation Project of Chinese Academy of Sciences.

**Figure captions**

FIG. 1. (Color online) KS (line with hollow squares) and their fits (solid line) with two symmetric Lorentz curves at a fixed time delay ($\tau_d$) of 30 ps, $T$=1.5 K and $B$=5.6 T.

FIG. 2. (Color online) Spin precessions with and without microwave irradiation at $E_p$=1.516 ev, $T$=1.5 K and $B$=5.6 T. (a), TRKR traces with (upper) and without (lower) microwave irradiation—data (symbol + line) and fits (solid line). (b), Corresponding FFTs of the experimental TRKR traces —with (hollow circles + line) and without (filled squares + line) microwave irradiation. Component resolving results of the TRKR traces without (c) and with (d) microwave.

FIG. 3. Decaying envelope of microwave-induced ΔTRKR measured at $E_p$=1.516ev/1.522ev at $T$=1.5 K, $B$=5.6 T. FFTs yield two different frequencies for each ΔTRKR trace, indicating that the microwave can affect two different spin precessions simultaneously.

**Figure 1**

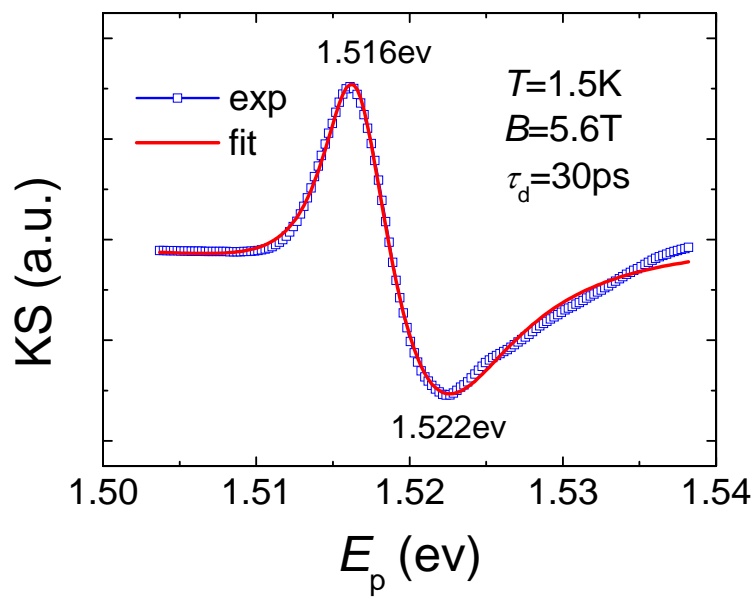

**Figure 2**

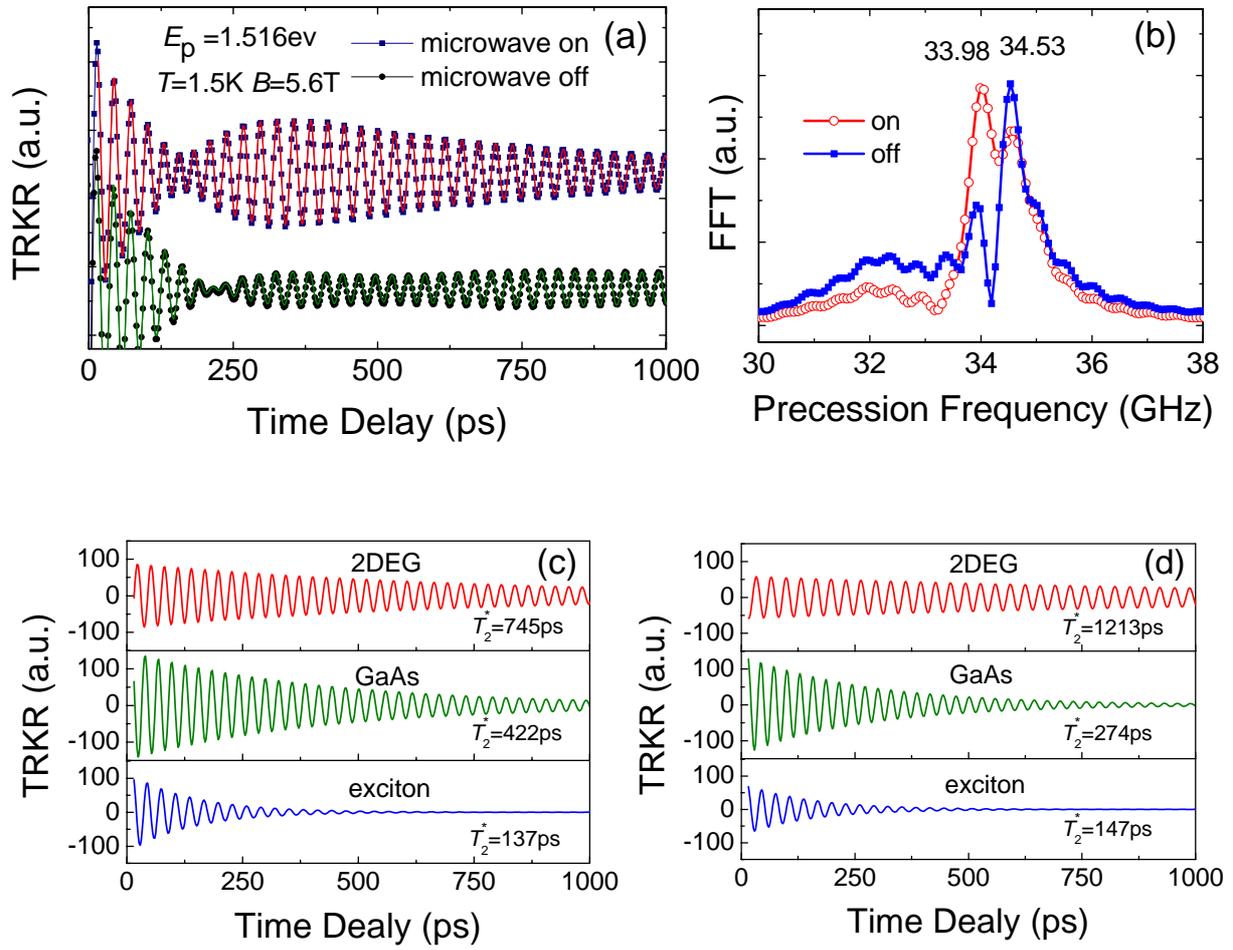

**Figure 3**

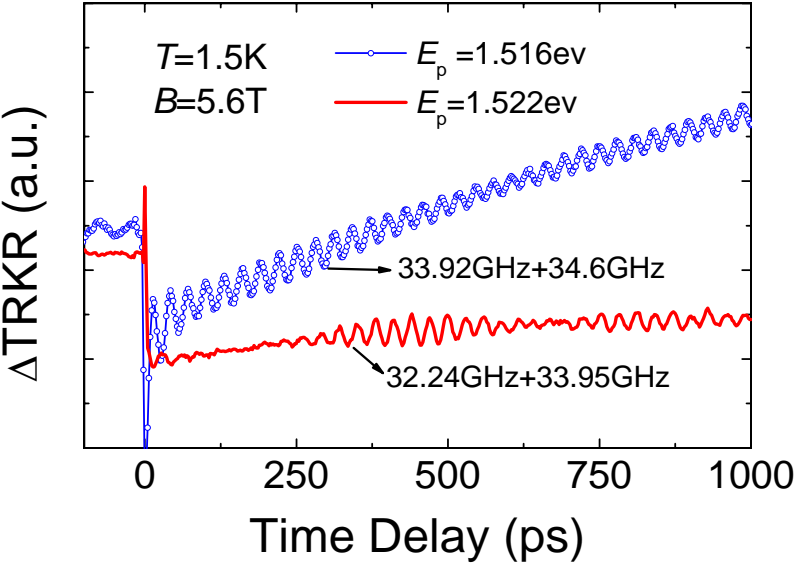